\documentclass[prd,aps,11pt,
nofootinbib,tightenlines,superscriptaddress,preprintnumbers]{revtex4}

\def\lsim{\raise0.3ex\hbox{$<$\kern-0.75em\raise-1.1ex\hbox{$\sim$}}}
\def\gsim{\raise0.3ex\hbox{$>$\kern-0.75em\raise-1.1ex\hbox{$\sim$}}}

 \usepackage{graphicx}

\newcommand{\beq}{\begin{equation}}
\newcommand{\eeq}{\end{equation}}
\newcommand{\bqa}{\begin{eqnarray}}
\newcommand{\eqa}{\end{eqnarray}}

\begin{document}

\title{\Large\bf Dissipative Hydrodynamics and Heavy Ion Collisions}

\preprint{BI-TP 2006/04}\preprint{RBRC 589}

\author{Rudolf Baier}

\author{Paul Romatschke}
\affiliation{Fakult\"at f\"ur Physik, Universit\"at Bielefeld, 
D-33501 Bielefeld, Germany}
\author{Urs Achim Wiedemann}
\affiliation{Department of Physics and Astronomy, University of Stony
Brook, NY 11794, USA}
\affiliation{RIKEN-BNL Research Center, Brookhaven National Laboratory, Upton,
NY 11973-5000, USA}
\date{\today}

\begin{abstract} 
Recent discussions of RHIC data emphasized the exciting possibility that the matter 
produced in nucleus-nucleus collisions shows properties of a near-perfect fluid. 
Here, we aim at delineating the applicability of fluid dynamics, which is needed to
quantify the size of corresponding dissipative effects. We start from the equations
for dissipative fluid dynamics, which we derive from kinetic theory up to second 
order (Israel-Stewart theory) in a systematic gradient expansion. In model studies, 
we then establish that for too early initialization of the
hydrodynamic evolution ($\tau_0 \lsim 1$ fm/c) 
or for too high transverse momentum 
($p_T \gsim 1$ GeV) in the final state, the
expected dissipative corrections are too large for a fluid description to be 
reliable. Moreover, viscosity-induced modifications of hadronic transverse momentum
spectra can be accommodated to a significant degree in an ideal fluid description by 
modifications of the decoupling stage. We argue that these conclusions, drawn from
model studies, can also be expected to arise in significantly
more complex, realistic fluid dynamics simulations of heavy ion collisions. 
\end{abstract}

\maketitle

\section{Introduction}

Why is it interesting to characterize dissipative effects of the dense QCD matter
produced in ultra-relativistic nucleus-nucleus collisions at RHIC or at the LHC?
The experimental heavy ion programs  aim at establishing properties of QCD
matter at the highest energy densities attained in the laboratory~\cite{Adcox:2004mh,Back:2004je,Arsene:2004fa,Adams:2005dq}. Shear viscosity,
bulk viscosity, heat conductivity or the conductivities of conserved charges are 
such properties of QCD matter. If unambiguously extracted from data, they
are prime candidates for the next compilation 
of the Particle Data Group. They are of fundamental interest, since they are not
mere material constants, but they are -- at least in principle -- computable from first
principles~\cite{Danielewicz:1984ww,Trans1,Trans2,AdS,visc}.

How can one extract dissipative transport coefficients from data? For this,
it is a prerequisite to have a dynamical theory of nucleus-nucleus collisions, which
can be compared to data and depends on dissipative properties. Dissipative 
hydrodynamics is this theory. It describes liquids, which deviate locally from
a fully equilibrated, 'perfect' one.  Dissipative hydrodynamics is not applicable to 
general non-equilibrium evolution. Deviations from ideal fluid dynamics must be 
sufficiently small for the gradient expansion underlying dissipative hydrodynamics 
to be valid. On the other hand, dissipative effects must be sufficiently large to be 
measurable. These two requirements limit the applicability of dissipative hydrodynamics 
to heavy ion collisions. They will complicate any attempt to determine transport 
coefficients from data.

Remarkably, simulations of Au-Au collisions at RHIC based on ideal fluid 
dynamics~\cite{Teaney:2000cw,Huovinen:2001cy,Kolb:2001qz,Hirano:2002ds,Kolb:2002ve} 
reproduce the observed large size and impact parameter dependence of elliptic 
flow for sufficiently central collisions ($b \leq 7$ fm) near 
mid-rapidity~\cite{Ackermann:2000tr,Adcox:2002ms,Adler:2003kt}. They
also account for the transverse momentum dependence of hadronic spectra up to
$p_T \leq 1.5$ GeV, and they reproduce the gross features of the particle species
dependence of these spectra~\cite{Adler:2003cb,Adams:2003xp}.
However, these ideal hydrodynamic descriptions of RHIC data
require a very short thermalization time, $\tau_0 < 1$ fm/c \cite{Heinz:2004pj}, for which multiple scattering 
models cannot account naturally in the weak coupling regime~\cite{Baier:2002bt,Molnar:2001ux}.
While this adds to the paradigm that high-energy heavy ion collisions 
create a strongly coupled plasma~\cite{Shuryak:2004cy,Lee:2005gw,Gyulassy:2004zy,Heinz:2005zg,Muller:2006ee,Blau:2005pk},
one cannot exclude a priori the existence of collective mechanisms (such as plasma 
instabilities~\cite{Mrowczynski:1996vh,Arnold:2004ti,Rebhan:2004ur,Dumitru:2005gp,Romatschke:2005pm}), which may 
account for fast equilibration 
in the weakly coupled regime. 

Ideal hydrodynamic simulations of Au-Au collisions at RHIC also gave support to the 
conclusion~\cite{Adcox:2004mh,Back:2004je,Arsene:2004fa,Adams:2005dq}
that for realistic initial conditions, the observed elliptic flow exhausts the theoretical
upper limit predicted by ideal fluid dynamics, thus indicating that a perfect liquid
with negligible dissipation has been created at RHIC~\cite{Shuryak:2004cy,Heinz:2005zg}. 
However, this statement must
be qualified, since recent work~\cite{Hirano:2005xf} has identified a class of a priori 
realistic initial conditions, for which ideal fluid dynamics significantly overestimates 
the elliptic flow measured at RHIC. 
Also, it has been questioned that the observed elliptic flow is indicative of full
equilibration~\cite{Bhalerao:2005mm}.
This illustrates the major limitation of hydrodynamic concepts to heavy ion 
collisions: while the dynamical description is parameter free, rather precise knowledge 
of the initial conditions is essential for the predictive or interpretational power of the
theory. This is so for ideal fluid dynamics already. But since the applicability of dissipative
fluid dynamics depends on precise knowledge about the closeness to local equilibrium,
one may expect that control over initial conditions becomes an even more sensitive
issue for the characterization of dissipative properties. 

The main motivation of this study is to delineate the region of applicability of (ideal as 
well as dissipative) fluid dynamics in heavy ion collisions. To this end, we consider 
dissipative hydrodynamics up to second order in the gradient expansion around ideal
hydrodynamics, the so-called Israel-Stewart theory~\cite{IS}. This theory has been applied to
viscous hydrodynamics only recently 
\cite{Muronga,Muronga:2004sf,Heinz:2005bw} (see however
\cite{Raju1,Raju2}), 
subjected to further
approximations which we discuss. 
The present work follows the 
standard perturbative logic, that to check the validity of 
the zeroth order (i.e. ideal hydrodynamics), one should check the size of higher orders
in the perturbative expansion (here: gradient expansion). An expansion to second order
is necessary, since first order dissipative hydrodynamics implicitly assumes vanishing
relaxation times and is known to show therefore acausal artifacts~\cite{Hisc}.
Beyond quantifying the 'closeness' of the time evolution to ideal hydrodynamics, the 
motivation for studying the gradient expansion of dissipative fluid dynamics is of course
to gain access to the fundamental dissipative properties of the produced 
matter, namely its transport coefficients. 

Our work is organized as follows: In Section~\ref{sec2}, we specify the
equations of motion of second order dissipative hydrodynamics. We have found
these equations in the literature 
subjected to further sometimes ad hoc approximations.
In Appendix \ref{appa}, we rederive these equations from Boltzmann transport
theory. This allows us to discuss explicitly differences between existing formulations.
We then specify to a simple model inspired by heavy ion collision in
Section \ref{sec3}, and we discuss analytical results for this model. In Section \ref{sec4} we
numerically solve the second order dissipative hydrodynamic equations.
This leads to a lower bound on the hydrodynamic initialization time $\tau_0$ and
upper bound $p_T^{\rm crit}$ for the transverse momentum up to which a fluid
description is reliable. We then proceed to numerically evaluate hadronic transverse
momentum spectra via the Cooper-Frye formalism~\cite{CooperFrye}, 
both for ideal and second-order dissipative hydrodynamics. These results illustrate the
extent to which particle spectra emitted from a dissipative fluid may be mimicked by an
ideal fluid dynamics description. We discuss our main conclusions in Section \ref{sec5}.

\section{Dissipative Hydrodynamics}
\label{sec2}

From the Boltzmann transport equations, one can derive the 
hydrodynamic equations of motion including dissipative corrections to 
second order (see Appendix \ref{appa},  \cite{IS})
\bqa
(\epsilon+p)D u^\mu&=&\nabla^\mu p-\Delta^\mu_{\nu} \nabla_\sigma
\Pi^{\nu \sigma}+\Pi^{\mu \nu} D u_\nu\, ,
\label{2.1}\\
D \epsilon &=& - (\epsilon+p) \nabla_\mu u^\mu+\frac{1}{2}\Pi^{\mu \nu}
\langle\nabla_\nu u_\mu\rangle\, ,
\label{2.2}\\
\tau_{\Pi} \Delta^\mu_\alpha \Delta^\nu_\beta D \Pi^{\alpha \beta}
+\Pi^{\mu \nu}&=&\eta  \langle\nabla^\mu u^\nu\rangle 
- 2 \tau_{\Pi} \Pi^{\alpha (\mu}\omega^{\nu)}_{\ \alpha}\, .
\label{2.3}
\eqa
Here, the energy density $\epsilon$ and the pressure $p$ are related by the
equation of state. The vector $u^\mu$ denotes the local fluid velocity, and
$\Pi^{\mu \nu}$ is the shear tensor. In the limit of vanishing relaxation time $\tau_\Pi$,
the shear tensor is given by instantaneous information about the gradients of the
fluid velocity, $\Pi^{\mu \nu}=\eta \langle\nabla^\mu u^\nu\rangle$. This is the definition
of shear viscosity in first order dissipative fluid dynamics. In principle, dissipative
corrections depend also on heat flow and bulk viscosity, as well as on the 
corresponding relaxation times. However, the effects of the latter 
are expected to be much smaller on general grounds. To arrive
at a transparent discussion, we shall neglect them in what follows. 
We use the following notation 
\bqa
d_\mu u^\nu &\equiv& \partial_\mu u^\nu+\Gamma_{\alpha \mu}^{\nu}
u^\alpha
\label{2.4}\\
D&\equiv&u_\mu d^\mu\label{2.5} \\
\nabla^\mu &\equiv& \Delta^{\mu \nu} d_\nu\label{2.6} \\
\Delta^{\mu\nu}&\equiv & g^{\mu \nu}-u^\mu u^\nu\label{2.7} \\
\omega^{\mu \nu}&=&\Delta^{\mu \alpha} \Delta^{\nu \beta} 
\frac{1}{2}\left(d_\beta u_\alpha-d_\alpha u_\beta\right)\label{2.8} \\
\langle A_\mu B_\nu\rangle &\equiv & A_\mu B_\nu+A_\nu B_\mu
-\frac{2}{3} \Delta_{\mu \nu} A_\alpha B^\alpha\label{2.9} \\
(A_\mu,B_\nu) &\equiv &
\frac{1}{2}\left(A_\mu B_\nu+A_\nu B_\mu\right)\label{2.10} \\
\left[A_\mu,B_\nu\right] &\equiv&
\frac{1}{2}\left(A_\mu B_\nu-A_\nu B_\mu\right)\, ,
\label{2.11}
\eqa
where $d_\mu$ is the covariant derivative, $\Gamma_{\alpha \beta}^\nu$
are the Christoffel symbols and $\omega^{\mu \nu}$ is the vorticity
tensor.

\subsection{Approximation to hydrodynamic equations}

In recent studies of second order dissipative fluid dynamics~\cite{Heinz:2005bw}, 
one uses for the equation of motion of the shear viscosity
\beq
\tau_{\Pi}D \Pi^{\mu \nu} +\Pi^{\mu \nu}= 
\eta \langle\nabla^\mu u^\nu\rangle\, .
\label{2.12}
\eeq
To understand how this is related to expression (\ref{2.3}),  we expand
Eq.(\ref{2.3}) in the form
\beq
\tau_{\Pi}\left[D \Pi^{\mu \nu}+\left(u^{\mu} \Pi^{\nu}_{\ \alpha}+
u^{\nu}\Pi^{\mu}_{\ \alpha}\right)(D u^\alpha)\right]+\Pi^{\mu \nu}
- \eta \langle\nabla^\mu u^\nu\rangle
  =-2\tau_\Pi \Pi^{\alpha (\mu}\omega^{\nu)}_{\ \alpha}\, .
 \label{2.13}
 \eeq
For (\ref{2.12}) and (\ref{2.13}) to agree, we require that 
$\Pi^{\alpha (\mu}\omega^{\nu)}_{\ \alpha}=0$
and $\Pi^{\mu \alpha} D u_\alpha = 0$. By construction, the shear tensor is orthogonal 
to the fluid velocity, $\Pi^{\mu \alpha} u_\alpha = 0$, and these two conditions can 
be written as
\begin{eqnarray}
         \omega^{\nu\mu} &=& 0\, ,
        \label{2.14}\\
                D u^\nu &=& 0\, .
         	\label{2.15}
\end{eqnarray}
Thus, to arrive at (\ref{2.12}), one has to {\it assume} that the liquid is vorticity free.
Moreover, one {\it assumes} ad hoc that $D u^\nu $ vanishes up to second order in
a gradient expansion. This second requirement is not satisfied for a general flow
field, but it holds for the simple example studied in Section \ref{sec3} below.

We note that for a consistent treatment, the assumptions (\ref{2.14})
and (\ref{2.15}) have to be applied to all three equations of motion 
(\ref{2.1})-(\ref{2.3}). 
The equations of motion (\ref{2.1})-(\ref{2.3}) are then replaced by Eq. (\ref{2.12}) and
\bqa
0 &=&\nabla^\mu p-\Delta^\mu_{\nu} \nabla_\sigma
\Pi^{\nu \sigma}\, ,
\label{2.16}\\
D \epsilon &=& - (\epsilon+p) \nabla_\mu u^\mu+\frac{1}{2}\Pi^{\mu \nu}
<\nabla_\nu u_\mu>\, .
\label{2.17}
\eqa

The equations of motion (\ref{2.1})-(\ref{2.3}) were obtained from transport theory.
An alternative derivation starts from the most general second order gradient
expansion of the entropy density $s$. Since entropy cannot decrease, this leads
to the requirement that the four-vector $s^{\mu}$ of the entropy density
~\cite{Rischke:1998fq} satisfies
\beq
T d_\mu s^\mu = \Pi_{\mu \nu}\left[-\beta_2 D \Pi^{\mu \nu}
+\frac{1}{2} <\nabla^\mu u^\nu>\right] \equiv \frac{1}{2 \eta}
\Pi_{\mu \nu} \Pi^{\mu \nu}\, ,
\label{2.18}
\eeq
where $\beta_2=\frac{\tau_\Pi}{2 \eta}$. From this relation it is
obvious that possible terms in the square bracket orthogonal to
$\Pi^{\mu \nu}$ cannot be found by this ``entropy-wise''
derivation of the hydrodynamic equations. As we have seen above, 
it is by these terms that Eq.(\ref{2.13}) differs from 
Eq.(\ref{2.12}).

\section{Ideal versus viscous fluid dynamics for a simplified model}
\label{sec3}

For general initial conditions, the hydrodynamic evolution according to 
Eqs. (\ref{2.1})-(\ref{2.3}) is 
complicated and requires a (3+1)-dimensional numerical simulation. Here, we 
consider a simplified model, which incorporates several characteristic features
of a heavy ion collisions, while lacking many of the technical complications
of the most general solution. In this Section, we specify the geometry, initial conditions,
equation of motion, equation of state and viscosity for this simple model.

\subsection{Geometry and initial conditions}
\label{sec3a}
For the discussion of relativistic heavy ion collisions, it is useful to transform
to radial coordinates in the transverse plane, and to proper time $\tau$ and
space-time rapidity $\eta$,
\beq
\tau=\sqrt{t^2-z^2}, \quad \eta={\rm arctanh}\frac{z}{t}, \quad 
r=\sqrt{x^2+y^2}, \quad \phi={\rm arctan}\frac{y}{x}\, .
	\label{3.1}
\eeq
The metric in these coordinates is diagonal, 
\begin{equation}
        g_{\mu\nu} = \left( g_{\tau\tau}, g_{rr}, g_{\phi\phi}, g_{\eta\eta}\right)
                = \left(1,-1,-r^2,-\tau^2\right)\, .
                \label{3.2}
\end{equation}
The only non-vanishing Christoffel symbols are
$\Gamma^{\tau}_{\eta
\eta}=\tau$, $\Gamma_{\tau \eta}^{\eta}=1/\tau$, 
$\Gamma_{\phi \phi}^{r}=-r$, $\Gamma_{r \phi }^{\phi}=1/r$.

To arrive at a simple model, we assume
that the initial conditions are longitudinally boost-invariant, i.e., 
the energy density has a vanishing gradient in $\eta$ and the
initial fluid velocity has the profile
\begin{equation}
 u^\mu=(u^\tau,u^r,u^\phi,u^\eta)=(1,0,0,0). 
 \label{3.3}
\end{equation}
The hydrodynamic evolution preserves this boost-invariance, which in ultrarelativistic
heavy ion collisions is expected to be realized near mid-rapidity~\cite{Bjorken}. 

Second, we assume that the initial conditions show neither flow nor density 
gradients in the transverse direction. This implies that the $r$- and $\phi$-dependence
of the hydrodynamic evolution becomes trivial.
This condition is realized for matter of $r$-independent density and infinite transverse 
extension. For realistic collision geometry, the edge of the transverse overlap region will 
always show sizeable transverse density gradients, which drive sizeable transverse 
flow gradients in subsequent steps of the dynamical evolution. However, for matter close to 
the transverse center of the collision, approximate $r$-independence of density and flow
can be realized for early times. 
This is also seen by multi-dimensional simulations 
of ideal fluid dynamics, which reveal for an extended time period approximately 
one-dimensional expansion dynamics. In this sense, we view the one-dimensional 
model resulting from the above-mentioned approximations as a model which retains 
characteristic features of a fully multi-dimensional simulation of heavy ion collisions. 

\subsection{Equation of state and viscosity}
\label{sec3b}

To close the hydrodynamic equations of motion, one has to specify an equation of
state. Here, we consider the ideal one,
\begin{equation}
	\epsilon = 3\, p\, . 
	\label{3.4}
\end{equation}
For numerical calculations, we shall use the perturbative result for the
pressure of the Yang-Mills theory to leading order in the 
coupling $g^2$
\begin{equation}
 p = \frac{\pi^2 T^4}{90}\left(2 (N_c-1)^2 + \frac{7 N_c
 N_f}{2}\right)\, .
 	\label{3.5}
\end{equation}
We consider QCD ($N_c=3$) with $N_f=0$ for simplicity, where
%
\beq
	a=p/T^4=\frac{8 \pi^2}{45}\, .
	\label{3.6}
\eeq
For numerical calculations, we also need to specify the dimensionless ratio
$\eta/s$. The perturbative result, given more explicitly in Appendix~\ref{appc},
is to leading logarithmic accuracy of the form
$\eta=\frac{T^3}{g^4} \frac{\eta_1}{\ln{\mu^{*}/m_D}}$. For QCD ($N_c=3$) and $N_f=0$, 
it takes the form~\cite{Trans1,Trans2} 
\beq
	\eta/s=\frac{1}{4a}\frac{27.126}{g^4
	\ln\left({2.765 g^{-1}}\right)}\, .
	\label{3.7}
\eeq
According to this expression, the range $\alpha_s=\frac{g^2}{4\pi}=(0.2,0.4)$)
corresponds to $\eta/s\simeq(1.1,0.73)$. However, often the logarithmic correction
in Eq.(\ref{3.7}) is assumed to be $\ln{\mu^{*}/m_D}\sim O(1)$. Then, one
finds for the same values of the coupling constant the range 
$\eta/s\simeq(0.61,0.15)$. This is one of many illustrations of the significant
numerical uncertainties related to the use of these results. On the other hand,
there is a conjecture~\cite{AdS}, that in all thermal gauge field theories
$\eta/s \geq \frac{1}{4\pi}\simeq 0.08$. This lower bound is realized in the strong
coupling limit of highly supersymmetric thermal Yang-Mills theories, and it
may be of relevance to the strong coupling regime of finite temperature QCD near
$T_c$. 

We emphasize that we do not advocate the applicability
of perturbative results to the dense QCD matter produced
in nucleus-nucleus collisions. The sole purpose of the above discussion
was to survey the range and
uncertainties of the numerical input, suggested by different perturbative and
non-perturbative calculational approaches.  

\subsection{Equations of motion}
\label{sec3c}
In our model, 
the $\eta$- and $r$-independence of initial density and flow profile implies strong
simplifications.  Making use of $\langle\nabla^\tau u^\tau\rangle=0$ and $\Pi^{\tau \tau}=0$
(following from $u_\alpha \langle\nabla^\alpha u^\beta\rangle =0=u_\alpha
\Pi^{\alpha \beta}$), and the trace condition $\Pi^{\mu}_\mu=0$, one finds after some algebra 
\beq
\frac{1}{2} \Pi^{\mu \nu} \langle\nabla_{\nu} u_{\mu}\rangle = - \tau \Pi^{\eta \eta}\, .
	\label{3.8}
\eeq
%
Since (\ref{2.14}) and (\ref{2.15}) are satisfied, the equations of motion take the form
\bqa
  0&=&\nabla^\mu p-\Delta^\mu_{\nu} \nabla_\sigma
  \Pi^{\nu \sigma}\, ,
  \label{3.9}\\
  D \epsilon &=& - (\epsilon+p) \nabla_\mu u^\mu-\tau \Pi^{\eta \eta}\, ,
 \label{3.10}\\
	\tau_{\Pi} D \Pi^{\eta \eta}
	+\Pi^{\eta \eta}&=&\eta \langle\nabla^\eta u^\eta\rangle\, .
   \label{3.11}
\eqa
Eq.(\ref{3.10}) can be simplified further by realizing that $\nabla_\mu u^\mu=1/\tau$.
Moreover, explicit calculation of (\ref{3.9}) results in 
\begin{equation}
	\nabla^\mu p = 0\, .
	\label{3.12}
\end{equation}
Thus, the present model describes matter with vanishing pressure gradients.
However, the model allows for a non-trivial time-dependent evolution of density 
since the matter is subject to Bjorken boost invariant expansion. 

We now discuss more explicitly, how leading and higher order dissipative effects enter the
equations of motion. In our model, all information about dissipative effects
resides in the component $\Pi^{\eta\eta}$ of the shear tensor.

\subsubsection{Zeroth order - Euler equations: the perfect liquid}
\label{sec3c1}
In the absence of dissipative effects, $\Pi^{\eta\eta}=0$, one recovers the equations of
motion of an ideal fluid
\beq
\partial_\tau \epsilon = - \frac{(\epsilon+p)}{\tau}\, .
	\label{3.13}
\eeq
For the equation of state $\epsilon=3 p$, this has the solution~\cite{Bjorken}
\beq
T(\tau)=T_0 \left(\frac{\tau_0}{\tau}\right)^{1/3}.
	\label{3.14}
\eeq
In what follows, this result defines the baseline on top of which dissipative
effects have to be established. 

\subsubsection{First order - Navier-Stokes: reheating artifacts at early times}
\label{sec3c2}
Dissipative corrections to first order in the gradient expansion are recovered by 
setting the relaxation time $\tau_\Pi=0$. This leads to
$\Pi^{\mu \nu}=\eta \langle\nabla^\mu u^\nu\rangle$, where $\eta$ is the (temperature
dependent) viscosity. The resulting equation of motion reads
\beq
	\partial_\tau \epsilon =-\frac{\epsilon+p}{\tau}+\frac{4 \eta}{3 \tau^2}\, .
	\label{3.15}
\eeq
For the equation of state and viscosity defined in Section~\ref{sec3b},
the solution of this differential equation is known analytically for the
case of constant $\eta/s$~\cite{Kouno,Muronga}
\beq
	T(\tau)=T_0 \left(\frac{\tau_0}{\tau}\right)^{1/3}\left[1+\frac{2\eta}{3 s
 \tau_0 T_0}\left(1-\left(\frac{\tau_0}{\tau}\right)^{2/3}\right)\right].
 \label{3.16}
\eeq
This function has a maximum temperature at time
\begin{equation}
	\tau_{\rm max} = \tau_0 \left( \frac{1}{3} + \frac{s}{\eta} \frac{\tau_0\, T_0}{2} \right)^{-3/2}\, .
	\label{3.17}
\end{equation}
%
For times $\tau > \tau_{\rm max}$, the temperature decreases with time, as expected
for matter undergoing expansion. For early times $\tau < \tau_{\rm max}$, however,
the solution shows an unphysical reheating. We note that first order dissipative fluid
dynamics is known to show unphysical effects~\cite{Hisc}. 

From a pragmatic point of view, one may ask how the numerical importance of
this reheating artifact depends on the time $\tau_0$, at which one initializes the 
dissipative evolution. For one-dimensional viscous first order hydrodynamics, the
entropy evolves according to $\frac{d(\tau s)}{d\tau} = \frac{4\eta/3}{\tau T}$. For a
perfect liquid, entropy does not change. As a consequence, for dissipative corrections
to be small, one requires that the increase of entropy is small compared to  
the total entropy in the system, i.e., $\frac{\eta}{s} \ll \tau\, T$ must
hold throughout the dynamical evolution. For initial time $\tau_0$ and initial
temperature $T_0$, we thus require $\frac{\eta}{s} \ll \tau_0\, T_0$, and we find
\begin{equation}
	\tau_{\rm max}\, T_0 = 
	\frac{1}{\sqrt{\tau_0\, T_0}} \left(  \frac{2\eta}{s}  \right)^{3/2}\, ,\qquad
		\hbox{\rm for} \qquad
	\frac{\eta}{s} \ll \tau_0\, T_0\, .
	\label{3.18}
\end{equation}
So, if $\frac{\eta}{s} \ll \tau_0\, T_0$, then $\tau_{\rm max} \ll \tau_0$ 
and one can expect that the unphysical reheating effect is not seen during the time studied
by the evolution. One may then hope that the calculation returns reasonable results, although
the absence of an obviously unphysical behavior by no means implies that. A check by
including higher order corrections appears to be advisable even in this range.
On the other hand, for fixed value
$\frac{\eta}{s} \ll 1$, one can always find sufficiently small times $\tau_0$ so that
the entropy density increase per unit time is large initially. For these early times,
dissipative effects will be large, and will lie outside the validity of the relativistic
Navier-Stokes equations. This is seen in particular in the appearance of an
unphysical reheating effect at early times.

\subsubsection{Second Order - Israel-Stewart theory: causal dissipative hydrodynamics}
\label{sec3c3}

For the boost-invariant model defined in Section \ref{sec3a}, the equations of motion to
second order in dissipative gradients read~\cite{Lallouet,Muronga}
\bqa
\partial_\tau \epsilon&=&- \frac{\epsilon+p}{\tau}+\frac{1}{\tau} \Phi\, ,
 \label{3.19}\\
 \partial_\tau \Phi &=&-\frac{\Phi}{\tau_\Pi}+\frac{2}{3 \tau \beta_2}\, ,
	\qquad \beta_2 \equiv \frac{\tau_\Pi}{2 \eta}\, .
	\label{3.20}
\eqa
Here, we have introduced
\begin{equation}
 \Phi \equiv - \tau^{2} \Pi^{\eta \eta}\, .
 \label{3.21}
 \end{equation}
The ratio $\beta_2$ of relaxation time over viscosity can be calculated 
for a massless Boltzmann gas by using Eq.(\ref{a.24}) and determining 
the pressure $p$. One finds
\begin{equation}
 \beta_2=\frac{3}{4 p}\, .
 \label{3.22}
\end{equation}
For a Bose-Einstein gas, this relation is only approximately true. It is modified by 
the factor $\sim1.024$ (see Appendix~\ref{appb}).
The equations have to be solved numerically which we do for 
the equation of state $\epsilon=3 p$. We will scan a range of  
initial conditions for the temperature and initial time 
$T_0,\tau_0$ as well as the strength of dissipative effects controlled
by $\eta/s$; for simplicity we always set $\Phi(\tau_0)=0$ in the 
following.

\section{Numerical results for second order viscous hydrodynamics}
\label{sec4}

\begin{figure}
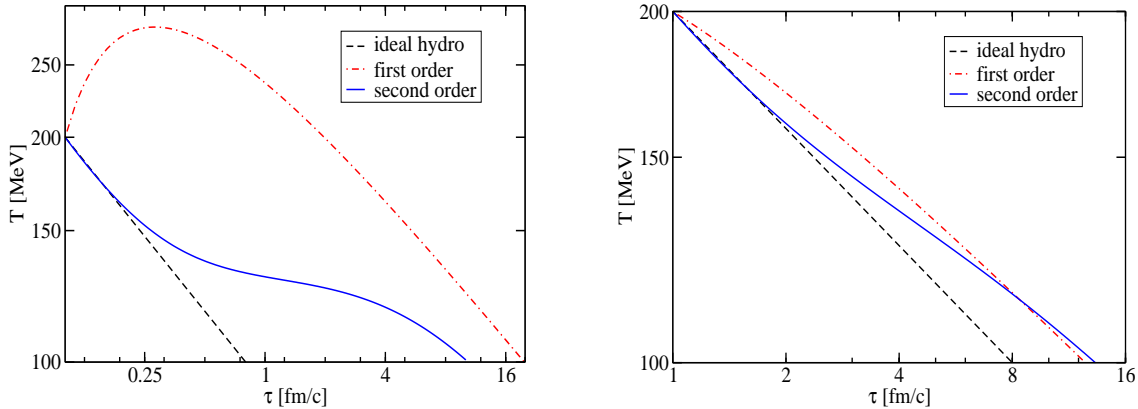

\vspace{0.6cm}
\begin{center}
\includegraphics[width=2.1in,height=2.7in,angle=-90]{fig1a.epsi}
\hspace*{1cm}
\includegraphics[width=2.1in,height=2.7in,angle=-90]{fig1b.epsi}
\end{center}
\caption{Temperature evolution in ideal hydrodynamics (dashed line), 
first order dissipative hydrodynamics (relativistic Navier-Stokes,
dash-dotted line) 
and second order dissipative hydrodynamics (Israel-Stewart, full
line), $\eta/s=0.3$. 
For sufficiently early initialization time $\tau_0$, the relativistic 
Navier-Stokes equation gives rise to an
increase in temperature and thus total system energy, which is not
seen in the Israel-Stewart theory (see text for details).}
\label{fig1}
\end{figure}

We now turn to the numerical solutions of the simplified hydrodynamic
model, defined above. We shall compare results for ideal
fluid dynamics to results in which viscous effects are treated to first and second
order. Our main motivation is to understand where hydrodynamics is reliable,
how its validity and predictive power depends on the initial conditions and which
traces are left by dissipative effects in the data. 
Fig.\ref{fig1} shows the time dependence of the temperature
for initial conditions $T_0=200$ MeV and $\tau_0=0.1$ fm/c (left plot)
compared to $\tau_0=1$ fm/c (right plot). Remarkably, for
early initial times, the first order (Navier-Stokes) equation results in an
unphysical increase of temperature or total system energy, which we discussed 
already in the context of Eq.~(\ref{3.16}). In contrast, the second order dissipative 
hydrodynamic equations lead to the monotonous decrease of temperature with
expansion time, which one expects on general grounds. In general, dissipative
effects prolong the cooling time. 
Initializing the system with the same initial temperature $T_0=200$ MeV at a later 
time $\tau_0 = 1$ fm/c leads to significantly smaller dissipative effects, see Fig.~\ref{fig1}.
In this case, the Navier-Stokes first order solution provides a quantitatively reasonable 
estimate for the causal second-order Israel-Stewart theory. This raises the question how the 
applicability of hydrodynamics depends on the initialization time $\tau_0$. 

\subsection{Breakdown of hydrodynamic picture at small $\tau_0$}
\label{sec4a}

\begin{figure}
\vspace{0.6cm}
\begin{center}
\includegraphics[width=2.5in,angle=-90]{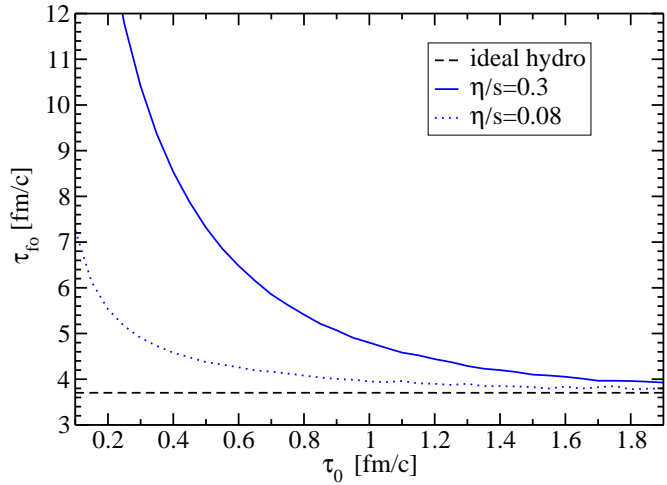}
\end{center}
\caption{The freeze-out time $\tau_{\rm fo}$ (at which the system
reaches $T_{\rm fo}=150$ MeV) as a function of initial time $\tau_0$ 
for ideal (dashed line) and second order dissipative hydrodynamics (solid line for 
$\eta/s=0.3$ and dotted line for $\eta/s=0.08$). The initial temperature $T_0$ was 
chosen such that $T_0^3\, \tau_0 = {\rm const}$, see text for details.}
\label{fig2}
\end{figure}

The two initial conditions used in Fig.\ref{fig1} are difficult to
compare, since the systems differ significantly in entropy density.
For the calculations
in Fig.~\ref{fig2}, we work with initial conditions for which 
$T_0^3\, \tau_0 = {\rm const}\sim \left( 251\, {\rm MeV}\right)^2 $.
For zeroth order ideal hydrodynamics, this ensures that the choice of the initialization
time $\tau_0$ does not affect the time evolution of the system, see  (\ref{3.14}). Irrespective
of $\tau_0$, the system has the same entropy, and thus leads to a final state with the same
event multiplicity. In this sense, the systems initialized at different $\tau_0$ but 
$T_0^3\, \tau_0 = {\rm const}$ are equivalent. 

However, source gradients are larger at earlier times. Therefore, dissipative fluid
dynamics will reveal stronger deviations from ideal fluid dynamics, if it is initialized
at smaller $\tau_0$. This is seen in Fig.~\ref{fig2}, where we plot the freeze-out time
$\tau_{\rm fo}$ (defined as the time at which the system reaches $T = 150$ MeV)
as a function of $\tau_0$. For ideal fluid dynamics, the freeze-out time is unaffected
by $\tau_0$. In the presence of dissipative effects, however, an earlier initialization
provides stronger gradients and more time for deviations from ideal hydrodynamics
to establish. In general, dissipative fluids take longer to cool. The larger the ratio
of viscosity over entropy density $\eta/s$, the more pronounced is the effect. 

If the gradients in the source are too large, then the gradient expansion underlying dissipative
hydrodynamics cannot be expected to converge. By estimating uncertainties
in 2nd order dissipative fluid dynamics, we find that this problem becomes 
relevant for $\tau_0 \lsim 1$ fm/c. At earlier times, dissipative hydrodynamics 
gradually loses its predictive and interpretational power.

\subsection{Breakdown of hydrodynamic picture at large $p_T$}
\label{sec4b}

The validity of a hydrodynamic picture is not only limited to sufficiently late times, but
also to sufficiently low transverse momenta. To estimate the maximum transverse
momentum $p_T^{\rm crit}$, up to which fluid dynamics may apply, we consider small
departures $\delta f\ll 1$ of the phase space distribution $f$ from its equilibrium value $f_0$.
For the gradient expansion underlying dissipative fluid dynamics to apply, one requires 
that local deviations $f-f_0$ of the phase space density are small compared to the phase
space density $f$. This leads to (see Appendix \ref{appa})
\beq
\delta f=\frac{1}{2 T^2(\epsilon+p)} \Pi_{\mu \nu} p^\mu p^\nu \ll 1.
\label{4.1}
\eeq
We now consider a particle at point $(\tau, \eta,r,\varphi)$, which follows the Bjorken
flow, $\eta = y$, and has transverse momentum $p_T$. Its four-momentum reads
$(p^t,p^x,p^y,p^z)=(m_T \cosh y,p_T \cos\varphi_p, p_T
\sin\varphi_p, m_T \sinh y)$, where $m_T^2=p_T^2+m_{\rm particle}^2$
depends on the particle mass $m_{\rm particle}$.
Averaging over the azimuthal angle $\varphi_p$, we find
that (\ref{4.1}) is equivalent to the condition 
\begin{equation}
	p_T^2\ll  \frac{4 T^2 (\epsilon+p)}{\Phi}\, .
	\label{4.2}
\end{equation}
Here, $\Phi$ determines the value of the shear tensor, see (\ref{3.21}). The dissipative
strength of the system is then quantified by the inverse Reynolds number
\cite{Baym}
\begin{equation}
	R_{\rm Reynolds}^{-1} = \frac{\Phi}{\epsilon+p}\, .
	\label{4.3}
\end{equation}
%

\begin{figure}
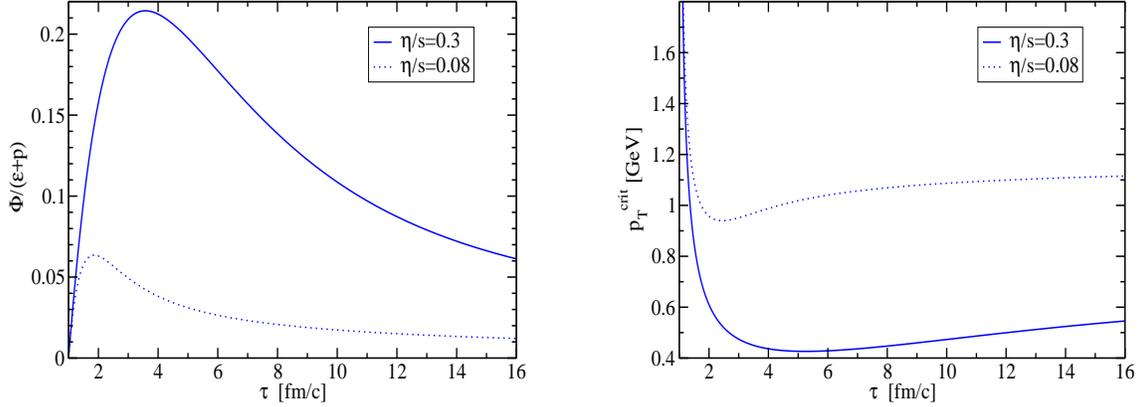

\vspace{.6cm}
\begin{center}
\includegraphics[width=2.1in,height=2.7in,angle=-90]{fig3a.epsi}
\hspace*{1cm}
\includegraphics[width=2.1in,height=2.7in,angle=-90]{fig3b.epsi}
\end{center}
\caption{
The time-dependence of the dissipative strength $\Phi/(\epsilon+p)$ (left plot) and 
the critical transverse momentum $p_T^{\rm crit}$ (right plot). Results are shown
for initial conditions $T_0=300$ MeV, $\tau_0=1$ fm/c, and for $\eta/s=0.3$ (solid line) 
and $\eta/s=0.08$ (dashed line), respectively.}
\vspace{0.6cm}
\label{fig3}
\end{figure}

In Figure \ref{fig3} we plot first a typical time evolution of the inverse Reynolds number
$\Phi/(\epsilon+p)$ for $T_0=300$ MeV, $\tau_0=1$ fm/c and different values $\eta/s=0.3$ 
and $\eta/s=0.08$. Remarkably, the value for $\Phi/(\epsilon+p)$ shows a maximum
at finite time. Therefore, in general, $p_T^{\rm
crit}=\sqrt{4 T^2 (\epsilon+p)/\Phi}$ assumes its lowest value before freeze-out. Thus, the maximum $p_T$ 
at which one can still trust a hydrodynamic description may actually be lower then the 
apparent value obtained at freeze-out. This is shown in the second part of
Fig.\ref{fig3}, where $p_T^{\rm crit}$ is plotted as a function
of time. As can be seen, for moderate values of $\eta/s\simeq 0.3$,
one obtains a $p_T^{\rm crit}< 0.5$ GeV.

\begin{figure}
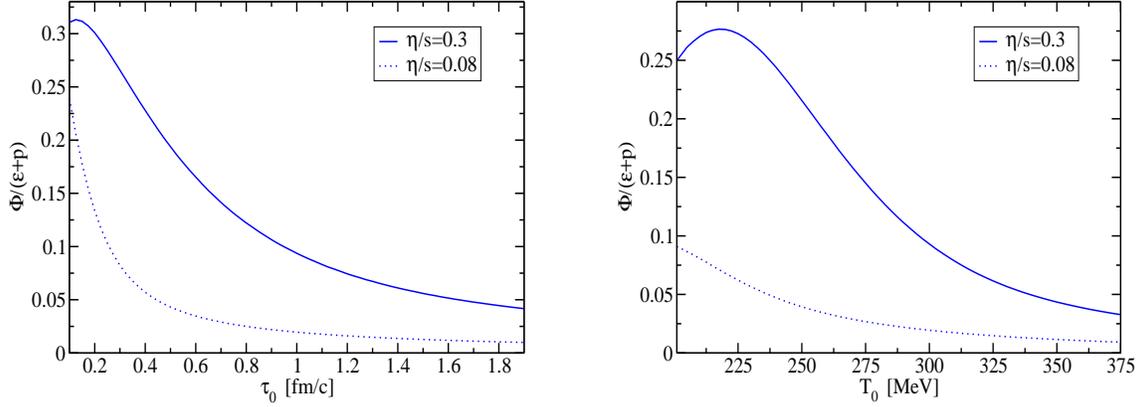

\vspace{0.6cm}
\begin{center}
\includegraphics[width=2.1in,height=2.7in,angle=-90]{fig4a.epsi}
\hspace*{1cm}
\includegraphics[width=2.1in,height=2.7in,angle=-90]{fig4b.epsi}
\end{center}
\caption{
The dissipative strength $\Phi/(\epsilon+p)$ at freeze-out ($T_{\rm fo}=150$ MeV)
as a function of initial time $\tau_0$ ($T_0=300$ MeV fixed, left
plot) and as a function of initial temperature $T_0$ ($\tau_0=1$fm/c
fixed, right plot) for $\eta/s=0.3$ (full lines) and $\eta/s=0.08$ (dotted lines).}
\label{fig4}
\end{figure}

In Fig.\ref{fig4}, we study how the inverse Reynolds number
$\Phi/(\epsilon+p)$ at the freeze-out temperature $T_{\rm fo}=150$ MeV depends on the
initial conditions $\tau_0$ and $T_0$. In general, the dissipative strength increases if
$\tau_0$ decreases or if $T_0$ decreases. 

\subsection{Hadron spectra at freeze-out}
\label{sec4c}

To calculate hadronic spectra from the phase space distribution $f$ evolved 
up to the freeze-out time $\tau_{\rm fo}$, one customarily makes use of the Cooper-Frye
freeze-out prescription~\cite{CooperFrye} 
\beq
	\frac{d^2 N}{d^2 p_T dy} = \int \frac{p_\mu
	d \Sigma^{\mu}}{(2\pi)^3} f\, .
	\label{5.1}
\eeq
Here, $d\Sigma_\mu$ denotes the oriented freeze-out volume. For the case of
a  Bjorken one-dimensional expansion, 
$p_\mu d\Sigma^\mu=m_T \cosh(y-\eta) \tau d\eta r dr d\phi$.
We now consider a phase space distribution $f$ at freeze-out, that deviates locally 
from equilibrium $f_0$. Adopting the ansatz of Ref.~\cite{Teaney}, which embeds
information about dissipative effects directly in the freeze-out distribution, we write 
\beq
	f=f_0 \left(1+\frac{1}{2 T^2 (\epsilon+p)} \Pi_{\mu \nu} p^\mu p^\nu\right)\, .
	\label{5.2}
\eeq
Since the relation (\ref{5.1}) between the particle spectrum and the phase space
distribution is linear, this allows us to define the particle spectrum in terms of its
ideal part and its dissipative corrections. Specializing to the case of a Boltzmann
distribution $f_0(p_\mu u^\mu)=2(N_c^2-1)\, \exp(-p_\mu u^\mu/T)$, one finds for the ideal
part\cite{Schnedermann}
\begin{eqnarray}
	\frac{d^2 N^{0}}{d^2 p_T dy} &=& \int \frac{p_\mu
	d \Sigma^{\mu}}{(2\pi)^3} f_0 \nonumber \\
	&=& 2(N_c^2-1) m_T \tau_{\rm fo}
	\frac{R_0^2}{(2 \pi)^2} K_1\left(\frac{m_T}{T}\right)\, .
\end{eqnarray}
The total yield is proportional to the transverse area $\pi\, R_0^2$ over which we
have integrated. For numerical calculations, we use $R_0 = 6$ fm.
For a Boltzmann distribution and longitudinally boost-invariant 
flow, the dissipative correction to this spectrum reads
\begin{eqnarray}
	\frac{d^2 \delta N}{d^2 p_T dy} &=& \frac{\Pi_{\alpha \beta}}{2 T^2 (\epsilon+p)} 
	\int \frac{p_\mu d \Sigma^{\mu}}{(2\pi)^3} f_0 p^\alpha p^\beta.
	\nonumber \\
	&=& \frac{\Phi}{4 (\epsilon+p)} 
	\frac{2(N_c^2-1) m_T \tau_{\rm fo} R_0^2}{(2 \pi)^2}
	 \left[\left(\frac{p_T}{T}\right)^2 K_1\left(\frac{m_T}{T}\right) 
	 \right.
	 \nonumber\\
	 && \qquad \qquad \left.
	-\frac{1}{2} \left(\frac{m_T}{T}\right)^2
	\left(K_3\left(\frac{m_T}{T}\right)-K_1\left(\frac{m_T}{T}
	\right)\right)\right].
\label{modspec}
\end{eqnarray}

\begin{figure}
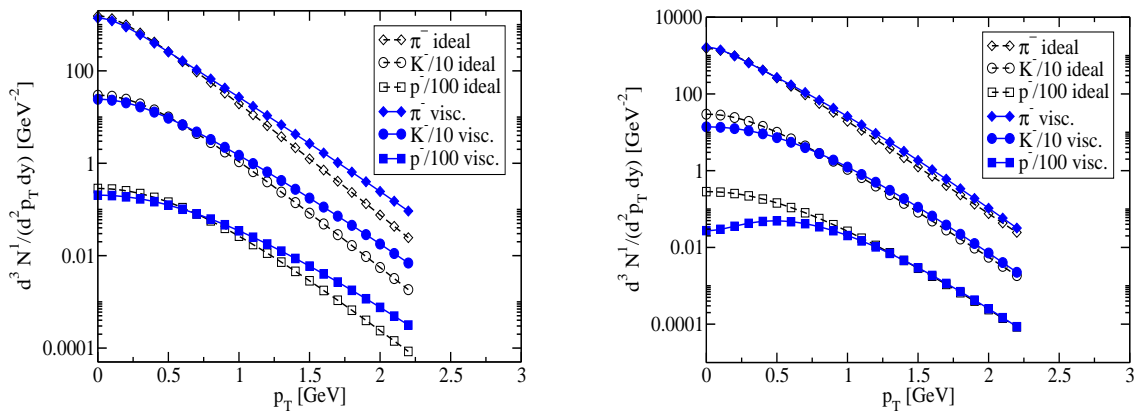

\vspace{0.6cm}
\begin{center}
\includegraphics[width=2.1in,height=2.7in,angle=-90]{fig5a.epsi}
\hspace*{1cm}
\includegraphics[width=2.1in,height=2.7in,angle=-90]{fig5b.epsi}
\end{center}
\caption{
Particle spectra for pions, kaons and protons from ideal
hydrodynamics (open symbols) and viscous hydrodynamics (full symbols),
for $R_0=6$ fm/c.
Left plot: spectra from ideal and viscous hydrodynamics ($\Phi/(\epsilon+p)=0.02$)
for the same freeze-out conditions.
Right plot: spectra for ideal and viscous hydrodynamics ($\Phi/(\epsilon+p)=0.05$)
with freeze-out of the ideal case adjusted such that slope of viscous calculation is
reproduced at intermediate $p_T$, (see text for details).}
\label{fig5}
\end{figure}

If $\eta/s$ is very small then dissipative corrections will be small. 
For instance, for the small value $\eta/s=0.08$, one finds a small dissipative
strength $\Phi/(\epsilon+p)\sim 0.02$ at freeze-out, if one uses $\tau_0>0.8$ fm/c 
and $T_0>300$ MeV for the initial conditions (see Fig.\ref{fig4}).
This estimate can be reconciled with the discussion in Ref.~\cite{Teaney},
if one uses $\tau_{\rm fo}\sim 9$ fm/c and $T_{\rm fo}\sim 150$ MeV so
that $\tau_{\rm fo} T_{\rm fo}\sim 7$ instead of $\tau_{\rm fo} T_{\rm fo}\sim
1$ used in \cite{Teaney}. For this small dissipative strength, we expect only 
slight modifications of hadron spectra. This is confirmed in Fig.\ref{fig5}.

However, freeze-out time $\tau_{\rm fo}$ and temperature $T_{\rm fo}$ are not directly
measurable. Rather, they are intrinsic parameters of the hydrodynamic model and its
possible matching to a hadronic rescattering phase.  This raises the question to what 
extent dissipative effects can be accommodated in an ideal fluid description of data 
by varying the freeze-out conditions  of the ideal fluid. In Fig.~\ref{fig5}, we have
calculated hadronic spectra $N^1=N^0+\delta N$ for a sizable dissipative strength $\Phi/(\epsilon+p)=0.05$ 
and freeze-out conditions $(\tau_{\rm fo},T_{\rm fo})= (14 {\rm fm/c},140 MeV)$.
Remarkably, these spectra can be reproduced to a significant degree by an 
ideal fluid model, using the same initialization but a different decoupling time and
temperature $(\tau_{\rm fo},T_{\rm fo})= (17 {\rm fm/c},170 MeV)$. 
The only clearly
visible difference between both models in Fig.~\ref{fig5} is a
characteristic mass-dependent 
dip in the low-$p_\perp$ part of the spectrum which increases with
increasing 
dissipative strength. This dip is a direct consequence of the analytical
form (\ref{modspec}),
and appears to be rather sensitive to small values of the inverse Reynolds
number. 
We have observed with curiosity that the fine-binned data available from 
BRAHMS~\cite{Ouerdane:2002gm} for $p_\perp > 0.5$ GeV for anti-protons
maybe hints a similar dip around the lower end of
the experimental 
reach. 
We regard this as an illustration that improved particle-identified
measurements
of hadronic transverse momentum spectra in the non-relativistic momentum
regime (e.g. smaller error bars from PHOBOS \cite{Wosiek:2002ur}) may 
provide valuable constraints for dissipative properties of the produced
dense matter.

\section{Conclusions}
\label{sec5}

The predictive and interpretational power of fluid dynamics simulations relies on 
knowledge about the initial conditions, the dissipative properties during evolution, 
and the modeling of decoupling in the final state, all of which are accompanied
by significant uncertainties in phenomenological applications. This raises the
question to what extent information about fundamental dissipative properties of dense
QCD matter can be constrained experimentally in heavy ion collisions. To address
this question, we have compared here results from ideal and dissipative fluid dynamics. 

First, from Boltzmann transport theory (see Section~\ref{sec2} and Appendices), we have 
established that derivations of second order hydrodynamics based on the implementation 
of $d_{\mu} s^{\mu} \geq 0$, (\ref{2.18})  do not provide the complete
set of second order terms in a gradient expansion. They discard those
terms in the gradient expansion which are orthogonal to the shear 
viscous tensor. 
We then explored in a model study the region of applicability of fluid
dynamics. We found that for too early initialization times 
$\tau_0 \lsim 1$ fm/c, source gradients are too large for dissipative hydrodynamics to 
converge (see Figs.~\ref{fig1} and ~\ref{fig2}). 
The inverse Reynolds number $\Phi/(\epsilon + p)$ defines a useful
$p_T$-dependent scale $p_T^{\rm crit}$ up to which dissipative 
hydrodynamics can be expected to account for deviations from the local equilibrium of 
a perfect  fluid.  For our model, we found typically a $p_T^{\rm crit} \lsim 1$ GeV. 

Once we had delineated for our model the range of validity of a fluid dynamics description,
we asked whether data taken within this range can be interpreted unambiguously in terms
of the strength of dissipative properties of matter. Our results illustrate the challenges
of this task. While even for a very small dissipative strength, model calculations of 
transverse hadron spectra show a measurable sensitivity to dissipative
properties, a very good control over initial
conditions and freeze-out is required 
to extract these features from data. 
In particular, in our model hadronic spectra including sizeable 
dissipative effects could be reproduced rather satisfactorily from perfect fluid calculations 
by modifying the freeze-out temperature and time (see Fig.~\ref{fig5}). 

The fluid dynamics model studied here is based on a strongly simplified transverse geometry
and does not account for a hadronic scattering phase. Therefore, 
this setup does not exhibit all characteristic features of a realistic heavy ion collision, such as 
transverse radial and elliptic flow. However, the two-step logic used here to assess the range of 
validity and the interpretational power of our model applies also to the more 
complex, realistic fluid dynamics simulations of heavy ion collisions: First, a comparison of
ideal and dissipative fluid dynamics results delineates the range of validity of a
fluid dynamics description. Second, within this limited range, one has to establish by 
model studies to what extent conclusions about the equation of state or dissipative
properties are independent of the initial conditions and the modeling of the freeze-out. 
For more complex models, there are more measurable quantities (such as elliptic
flow), which may help to constrain the fluid dynamics. However, there are also more 
features of the initial conditions, which require specification
(such as the initial transverse density and flow profile). This increases the complexity of
the task to disentangle effects of initial conditions from those of the properties of matter
determining fluid dynamics behavior. We expect that the numerical values for the bounds 
on initialization time and transverse momentum of a reliable fluid description, established here, 
change only weakly by going to more complex fluid simulations. 

\acknowledgments

PR was supported by BMBF 06BI102.
UAW acknowledges the support of the Department of Physics and Astronomy,
University of Stony Brook, and of RIKEN, Brookhaven National Laboratory, and the 
U.S. Department of Energy [DE-AC02-98CH10886] for providing the facilities essential 
for the completion of this work. 

\appendix
\section{Dissipative hydrodynamics from kinetic theory: Derivation for a massless
Boltzmann gas}
\label{appa}

In this Appendix, we derive the equation of motion (\ref{2.3}) of dissipative hydrodynamics
from kinetic theory for a massless Boltzmann gas. Our starting point is the 
Boltzmann transport equation
\beq
	p^\mu d_{\mu} f({\bf x},t,{\bf p}) = {\mathcal C}(x)\, ,
	\label{a.1}
\eeq
with ${\mathcal C}$ the collision term. We make use of the notational shorthands
(\ref{2.4})-(\ref{2.11}). We consider the first three moments of the Boltzmann
equation (\ref{a.1}), using $\int d\omega \equiv \int \frac{d^3 p}{(2\pi)^3\, p_0}$ 
\bqa
	\int\ d\omega p^\mu d_\mu f = d_\mu N^\mu = &0& = \int d\omega\
	{\mathcal C}\, ,
	\label{a.2}\\
	\int d\omega\ p^\mu p^\alpha d_\mu f = d_\mu T^{\mu \alpha} = &0& = 
	\int d\omega\ p^\alpha {\mathcal C}\, ,
	\label{a.3} \\
	\int d\omega\ p^\mu p^{\alpha} p^{\beta} d_\mu f &=& \int d\omega\
	p^{\alpha} p^{\beta}
	{\mathcal C}\, .
	\label{a.4}
\eqa
Here, equations (\ref{a.2}) and (\ref{a.3}) stand for charge conservation and
energy-momentum conservation, respectively. Equation (\ref{a.4}) is the lowest
order relation of moments of (\ref{a.1}), which contains dynamic information.
For simplicity, we consider in the following an equilibrium distribution $f_0$ of
the Boltzmann type for one degree of freedom, $f_0(u_\mu p^\mu)=\exp(-\beta u_\mu
p^\mu)$. Also, we employ the relaxation time approximation in Eq.(\ref{a.4}),
\beq
	{\mathcal C}=-p_\mu u^\mu \frac{f-f_{0}}{\tau_{\Pi}}\, .
	\label{a.5}
\eeq
For all points $({\bf x},t,{\bf p})$ in phase space, where we want a 
hydrodynamic description to apply, we assume that departures from 
local equilibrium are small, 
\beq
	f=f_0 (1+\delta f)\, . \qquad \vert\delta f\vert\ll1
	\label{a.6}
\eeq
The correction term $\delta f$ can be written  in the form
\beq
\delta f({\bf x},t,{\bf p})=\epsilon({\bf x},t) 
+ \epsilon_{\lambda}({\bf x},t) p^\lambda+
\epsilon_{\lambda \nu}({\bf x},t) p^\lambda p^\nu.
\label{a.7}
\eeq
The deviations from the equilibrium energy-momentum tensor $T^{\mu \nu}_0$ can 
be likewise written as
\beq
T^{\mu \nu}=T^{\mu \nu}_0+\Pi^{\mu \nu}+\Pi \Delta^{\mu \nu}
+q^{(\mu} u^{\nu)}\, . 
\label{a.8}
\eeq
Here, the deviations from local equilibrium due to shear viscosity,
bulk viscosity and heat flow are parametrized by $\Pi^{\mu \nu}$,
$\Pi$ and $q^{\mu}$, respectively. We limit the following discussion to
the effects of shear viscosity, which is expected to provide for the
most significant dissipative effects. 
It can be shown \cite{IS} that in this case we may ignore
both the zeroth and first order term in $p$ in Eq.(\ref{a.7})
as well as concentrate on $T^{<\mu \nu>}$ only.
We are left with $\delta f({\bf x},t,{\bf p})= \epsilon_{\lambda
\nu}({\bf x},t) p^\lambda p^\nu $, so that the second moment of the 
Boltzmann equation reads
\beq
d_\mu \int d\omega p^\mu p^{<\alpha} p^{\beta>} f_{0}\left(1+
\epsilon_{\gamma \delta} p^\gamma p^\delta\right)
 = - 
\frac{\epsilon_{\gamma \delta} u_\epsilon}{\tau_{\Pi}} \int d\omega
p^{<\alpha} p^{\beta>} p^\gamma p^\delta p^\epsilon f_{0}\, .
\label{a.9}
\eeq
To derive from this expression the equation of motion for the non-equilibrium
deviation $\Pi_{\mu\nu}$ of the energy momentum tensor, we establish first 
the relation between $\epsilon_{\mu\nu}$ and $\Pi_{\mu \nu}$. We start from
\bqa
<T^{\mu \nu}-T^{\mu \nu}_0> &=&
	\int d\omega p^{<\mu} p^{\nu>} (f-f_0) 
	\nonumber \\
	&=&  \epsilon_{\alpha \beta} (x)
	\int d\omega p^{<\mu} p^{\nu>} p^{\alpha} p^{\beta} f_{0}
	\nonumber \\
	&=& \epsilon_{\alpha \beta} (x) I^{<\mu \nu> \alpha \beta}\, ,
\label{a.10}
\eqa
where 
\beq
	I^{\mu_{1} \mu_{2}\ldots \mu_{n}} \equiv \int d\omega p^{\mu_1} p^{\mu_2}
	\ldots p^{\mu_n} f_0\, .
	\label{a.11}
\eeq
For the Boltzmann distribution $f_0=\exp{(- p_\mu u^\mu/T)}$, the integral (\ref{a.11})
can be solved analytically for general $n$ \cite{IS}. In the following, we need the
cases $n=4,5$ which read
\bqa
I^{\mu \nu \alpha \beta} &=& a_0 u^\mu u^\nu u^\alpha u^\beta 
	+a_1\left( \Delta^{\mu \nu} u^\alpha u^\beta + perm\right)
	\nonumber \\
	&& +a_2 \left(\Delta^{\mu \nu} \Delta^{\alpha \beta}+
	\Delta^{\mu \alpha} \Delta^{\nu \beta}+\Delta^{\mu \beta}
	\Delta^{\alpha \nu}\right)\, ,
	\label{a.12}\\
	I^{\mu \nu \alpha \beta \gamma}&=&b_0 u^\mu u^\nu u^\alpha u^\beta
	u^\gamma +b_1 \left(\Delta^{\mu \nu} u^\alpha u^\beta u^\gamma+perm.\right)
	\nonumber\\
	&&+b_2 \left(\Delta^{\mu \nu} \Delta^{\alpha \beta} u^\gamma + perm.\right)\, .
	\label{a.13}
\eqa
The temperature dependent coefficients $a_i$ and $b_i$ in (\ref{a.12}) and (\ref{a.13})
can be calculated e.g. by going to the fluid rest-frame, $u^\mu=(1,{\bf 0})$. One finds
$a_2=4 T^6/\pi^2$ and $b_2=24 T^7/\pi^2$. The first of these coefficients is related
to the pressure $p$ and energy density $\epsilon$ as $a_2=T^2(\epsilon+p)$. 
Combining Eqs. (\ref{a.8}), (\ref{a.10}) and (\ref{a.12}), we find
\bqa
	\epsilon_{\alpha \beta} (x) I^{<\mu \nu> \alpha \beta} 
	&=&
	2 a_2 \epsilon_{\alpha \beta} \Delta^{\alpha < \mu} \Delta^{\nu> \beta}
	\nonumber \\
	&=& 2 \Pi^{\mu \nu} =
	\Pi^{<\mu \nu>}\, .
	\label{a.14}
\eqa
This equation implies $u_\mu \Pi^{\mu \nu}=0$
and $\Delta_{\mu \nu} \Pi^{\mu \nu}=0$. Therefore, we can set
\begin{equation}
	\epsilon^{\mu \nu}=\frac{1}{2 a_2} \Pi^{\mu \nu}\, .
	\label{a.15}
\end{equation}
This allows us to
rewrite Eq.(\ref{a.9}) as
\beq
d_{\mu} \left(I^{\mu <\alpha \beta>}+\epsilon_{\gamma \delta} 
I^{\mu <\alpha \beta> \gamma \delta} \right)= - 
\frac{\epsilon_{\gamma \delta} u_\epsilon}{\tau_{\Pi}}
I^{<\alpha \beta> \gamma \delta \epsilon}
\label{a.16}
\eeq
The three terms entering this equation can be rewritten  with the help of the
identities derived above,
\bqa
&&\epsilon_{\gamma \delta} u_\epsilon I^{<\alpha \beta> \gamma \delta \epsilon}
	=2b_2 \epsilon_{\gamma \delta} \Delta^{\gamma <\alpha}
	\Delta^{\beta> \delta}=\frac{b_2}{a_2} \Pi^{<\alpha \beta>}\, ,
	\label{a.17}\\
	&&d_\mu \int d\omega p^\mu p^{<\alpha} p^{\beta>} f_{0} =
	- I^{\mu <\alpha \beta> \gamma} d_\mu \frac{u_\gamma}{T} = 
	-\frac{2 a_2}{T} <\nabla^{\alpha} u^\beta>\, ,
	\label{a.18}\\
	&&I^{\mu <\alpha \beta> \gamma \delta} d_{\mu} \epsilon_{\gamma \delta} 
	= 2 b_2 \left( \Delta^{\gamma < \alpha} \Delta^{\beta> \delta} 
	u^\mu d_\mu\epsilon_{\gamma \delta}+2 u^\gamma \Delta^{\mu <\alpha} \Delta^{\beta >
	\delta} d_\mu \epsilon_{\gamma \delta}\right)\, .
	\label{a.19}
\eqa
Further simplifications arise from
$u^\gamma d_{\mu} \Pi_{\gamma \delta}=-\Pi_{\gamma \delta} d_\mu
u^\gamma$ and $\Delta_{\gamma \delta} d_\mu \Pi^{\gamma \delta}=
d_\mu \Delta_{\gamma \delta} \Pi^{\gamma \delta} =0$. We then decompose
$\Pi^{\mu \nu} d_\mu u_\nu$  into a totally symmetric  component $\propto \theta_{\mu \nu}$,
and a totally antisymmetric expression proportional to the vorticity tensor $\omega_{\nu
\mu}$
\bqa
\Pi^{\gamma \delta} d_\mu u_\gamma &=& \Pi^{\gamma \delta} (\omega_{\gamma
	\mu}+\theta_{\gamma \mu})\, ,
	\label{a.20}\\
\omega^{\mu \nu}&=&\Delta^{\mu \alpha} \Delta^{\nu \beta} d_{[\beta}
	u_{\alpha]}\, ,
	\label{a.21}\\
\theta^{\mu \nu}&=&\Delta^{\mu \alpha} \Delta^{\nu \beta} d_{(\beta}
	u_{\alpha)}\, .
\label{a.22}
\eqa
At this point, we follow \cite{IS} in assuming the 'rigidity of flow'. This means
that one assumes the equilibrium flow $u^\mu$ to be shear free, and thus 
terms proportional to $\theta^{\mu \nu}$ to vanish. As a consequence, the term
$\epsilon_{\gamma \delta} 
\int d\omega p^\mu p^{<\alpha} p^{\beta>} p^{\gamma} p^{\delta} d_\mu
f_{0}$ can be neglected. Then, Eq.(\ref{a.16})  reads
\beq
\frac{-a_2}{T} <\nabla^\alpha u^\beta>+
\frac{b_2}{a_2} \left( \Delta^{\gamma \alpha} \Delta^{\beta \delta} 
D \Pi_{\gamma \delta}+2 \Pi^{\delta (\alpha}
\omega^{\beta)}_{\ \delta} \right)
+b_2 \Pi^{\alpha \beta} D a_2^{-1} =-\frac{1}{\tau_{\Pi}}
\frac{b_2}{a_2} \Pi^{\alpha \beta}.
\label{a.23}
\eeq
For the third term on the left hand side, $D a_2^{-1}\propto D \epsilon
\propto d_\mu u^\mu \propto \theta_\mu^\mu$, which vanishes under the 
assumption of rigid flow. We note that in Ref.~\cite{Muronga, Muronga:2004sf}, 
this term is kept. The contractions
$u_\nu d_\mu T^{\mu \nu}=0$ and $\Delta_{\nu \alpha} d_\mu T^{\mu \nu}=0$
for an energy momentum tensor of the form $T^{\mu \nu}=T^{\mu \nu}_0+\Pi^{\mu \nu}$
lead to the equations of motion (\ref{2.1}) and (\ref{2.2}), respectively.
To complete the derivation of (\ref{2.3}), we require that in the limit 
$\tau_\Pi \to 0$, the first order dissipative Navier-Stokes equations 
are recovered from (\ref{a.23}). This leads to the identification
\beq
\frac{a_2^2}{T b_2}\rightarrow \eta/\tau_{\Pi}\, .
\label{a.24}
\eeq
The equation of motion (\ref{2.3}) then follows from (\ref{a.23}).

\section{Dissipative Hydrodynamics: Derivation for a massless Bose-Einstein gas and 
momentum-dependent relaxation time}
\label{appb}

In this Appendix we derive the second order hydrodynamic equation (\ref{2.3}) for the 
case that the relaxation time $\tau_\Pi$ is momentum-dependent, $\tau_\Pi=p_\mu
u^\mu \hat{\tau}$, $\hat{\tau}={\rm const}$. This ansatz is of interest since it often enters
calculations of the shear viscosity (see Refs.~\cite{visc,relax,Heiselberg} and sec. 8.3
of the textbook~\cite{LeBellac}). In contrast to Appendix~\ref{appa}, we start here from
the equilibrium distribution for a massless Bose-Einstein gas $f_0=[\exp(\beta u_\mu
p^\mu)-1]^{-1}$, having in mind QCD with gluonic degrees of freedom.
This implies that the departures from equilibrium are expressed by
[c.f. Eq.(\ref{a.6})]
\beq
\delta f({\bf x},t,{\bf p})=(1+f_0) \epsilon_{\lambda \nu}({\bf x},t)
p^\lambda p^\nu\, .
\label{b.1}
\eeq
The relation between $\epsilon_{\mu \nu}$ and $\Pi_{\mu \nu}$ is
expressed as in Eq.(\ref{a.15}), except that the coefficient $a_2$ is
replaced by $\hat{a}_2 = 4 \zeta(5) T^6/\pi^2$, with $\zeta(n)$ the
Riemann $\zeta$-function.
This implies that $\hat{a}_2=\frac{\zeta(5)}{\zeta(4)} T^2
(\epsilon+p)$ when expressed in terms of energy density and pressure
of the free Bose-Einstein gas.
This modification, compared to the Boltzmann gas used in
App.\ref{appa}, is due to the replacement of the integrals
$I^{\mu_1 \mu_2\ldots \mu_n}$ by
\beq
{\hat I}^{\mu_1 \mu_2\ldots \mu_n} = \int d\omega
p^{\mu_1} p^{\mu_2}\ldots p^{\mu_n} f_0 (1+f_0)\, .
\label{b.2}
\eeq
The second moment of the Boltzmann equation (\ref{a.1}) reads now
\beq
d_{\mu}\left(I^{\mu<\alpha \beta>}+\epsilon_{\gamma \delta} 
\hat{I}^{\mu <\alpha \beta> \gamma \delta}\right) =
-\frac{\epsilon_{\gamma \delta}}{\hat{\tau}} \hat{I}^{<\alpha \beta>
\gamma \delta}\, ,
\label{b.3}
\eeq
where the right hand side depends on an order four tensor only, in contrast
to the order five tensor entering Eq.(\ref{a.16}).
Consequently, for a Bose-Einstein gas we find
\bqa
d_\mu I^{\mu \langle\alpha \beta\rangle}&=&-\frac{2 \hat{a}_2}{T} 
	\langle\nabla^\alpha u^\beta\rangle  \, ,
	\label{b.4}\\
\epsilon_{\gamma \delta} \hat{I}^{<\alpha \beta> \gamma \delta} &=&
	\Pi_{\gamma \delta} \Delta^{\gamma <\alpha} \Delta^{\beta > \delta}
	= 2 \Pi^{\alpha \beta}\, ,
	\label{b.5}\\
\hat{I}^{\mu <\alpha \beta> \gamma \delta} d_{\mu} \epsilon_{\gamma \delta} 
 	&=& 2 \hat{b}_2 \left( \Delta^{\gamma < \alpha}
	\Delta^{\beta> \delta} u^\mu d_\mu\epsilon_{\gamma \delta}
	+2 u^\gamma \Delta^{\mu <\alpha} \Delta^{\beta >
	\delta} d_\mu \epsilon_{\gamma \delta}\right)\, ,
	\label{b.6}
\eqa
where
\beq
	\hat{b}_2=\frac{24}{\pi^2} \zeta(6) T^7\, .
	\label{b.7}
\eeq
Together with the assumption of the ``rigidity of flow'', i.e. setting
$\theta_{\mu\nu} \equiv 0$, Eq.(\ref{b.3}) becomes
\beq
\frac{-\hat{a}_2}{T} <\nabla^\alpha u^\beta>+
\frac{\hat{b}_2}{\hat{a}_2} 
\left( \Delta^{\gamma \alpha} \Delta^{\beta \delta} 
D \Pi_{\gamma \delta}+2 \Pi^{\delta (\alpha}
\omega^{\beta)}_{\ \delta} \right)=-\frac{1}{\hat{\tau}}
\frac{\hat{b}_2}{\hat{a}_2} \Pi^{\alpha \beta}.
\label{b.8}
\eeq
Thus, we recover Eq.~(\ref{2.3}) upon identifying 
\beq
\eta \leftrightarrow \frac{\hat{a}_2}{T} 
\hat{\tau},\quad \tau_\Pi \leftrightarrow \frac{\hat{b}_2}{\hat{a}_2}
\hat{\tau},
\label{b.9}
\eeq
which leads to the ratio 
\beq
\beta_2 = \frac{\hat{b}_2 T}{2 (\hat{a}_2)^2} = \frac{3}{4 p}
\frac{\zeta(4) \zeta(6)}{\zeta(5)^2} \simeq 1.024 \frac{3}{4 p}\, .
\label{b.10}
\eeq
In conclusion, the dissipative hydrodynamic equations for a massless
Boltzmann and Bose-Einstein gas differ, but only on the order of a few percent.

\section{Perturbative calculation of the relaxation time}
\label{appc}

To complete the discussion on viscosity and relaxation time, we
calculate in QCD the relaxation time $\hat{\tau}=\tau_\Pi (p_\mu
u^\mu)^{-1}$ to leading order in the strong coupling constant
$g$. We follow the line of argument of 
Refs.\cite{visc,Heiselberg,LeBellac} for a gluonic system with $2 (N_c^2-1)=16$ degrees
of freedom. The collision term $\mathcal C$, entering this calculation,
is then dominated by $2\leftrightarrow2$ gluon scattering processes. 
The corresponding effective matrix element $M_{gg}$ is given in the 
hard-thermal loop (HTL) approximation, including Landau damping~\cite{LeBellac}.
To simplify the calculation, a small time-independent velocity $u^x(y)$
is considered which only varies with the space coordinate $y$~\cite{visc}, 
\beq
p_\mu d^\mu f({\bf x},t,{\bf p})\simeq - f_0(1+f_0) \frac{p^x p^y}{T}
\beta \frac{\partial u^x(y)}{\partial y}\, .
\label{c.1}
\eeq
Following analogous steps as in \cite{visc,LeBellac}, 
$\hat{\tau}$ can be written in terms of the ratio
\bqa
\frac{1}{\hat{\tau}}&=&\frac{2 (N_c^2-1)}{16} \left[
	\prod_{i=1}^4 \int d\omega_i (2\pi)^4 \delta^4(p_1+p_2-p_3-p_4)
	\overline{|M_{gg}|^2}(p_1^x p_1^y)^2 f_0(p_1)  \right.
	\nonumber \\
&&\left.
 	\times f_0(p_2) \left(1+f_0(p_3)\right)
	\left(1+f_0(p_4)\right)\right]/\left(\int d\omega (p^x p^y)^2 f_0 (1+f_0)\right)\, .
	\label{c.2}
\eqa
Here, $\overline{|M_{gg}|^2}$ denotes the scattering matrix element
squared, summed (averaged) over spins and color degrees of freedom in
the final (initial) state. The integrals are evaluated e.g. in \cite{LeBellac}, leading to
\beq
\frac{1}{\hat{\tau}} = \frac{(N_c^2-1) \pi^5 T^2 }{120\ \zeta(5)} 
\frac{g^4}{(4\pi)^2} \ln(4 \pi/g^2)\, .
\label{c.3}
\eeq
From Eq.(\ref{b.9}), the estimate for $\eta$ is derived
\beq
\eta = \frac{\hat{a}_2}{T} \hat{\tau} = \frac{960}{\pi^7}
	\zeta^2(5) T^3 \frac{(4\pi)^2}{g^4 \ln(4 \pi/g^2)},
\eeq
when taking the color degrees of freedom into account. We note that
$\hat{a}_2$ is proportional to a factor $2 (N_c^2-1)$. 
This expression agrees also with the value quoted in \cite{Trans1}.



\begin{thebibliography}{99} 
%
\bibitem{Adcox:2004mh}
  K.~Adcox {\it et al.}  [PHENIX Collaboration],
  Nucl.\ Phys.\ A {\bf 757} (2005) 184.
%
\bibitem{Back:2004je}
  B.~B.~Back {\it et al.} [PHOBOS Collaboration],
  Nucl.\ Phys.\ A {\bf 757} (2005) 28.
%
\bibitem{Arsene:2004fa}
  I.~Arsene {\it et al.}  [BRAHMS Collaboration],
  Nucl.\ Phys.\ A {\bf 757} (2005) 1.
%
\bibitem{Adams:2005dq}
  J.~Adams {\it et al.}  [STAR Collaboration],
  Nucl.\ Phys.\ A {\bf 757} (2005) 102.

\bibitem{Danielewicz:1984ww}
  P.~Danielewicz and M.~Gyulassy,
  Phys.\ Rev.\ D {\bf 31} (1985) 53.

\bibitem{Trans1}
  P.~Arnold, G.~D.~Moore and L.~G.~Yaffe,
  JHEP {\bf 0011} (2000) 001.

\bibitem{Trans2}
  P.~Arnold, G.~D.~Moore and L.~G.~Yaffe,
  JHEP {\bf 0305} (2003) 051.

\bibitem{AdS}
  G.~Policastro, D.~T.~Son and A.~O.~Starinets,
  Phys.\ Rev.\ Lett.\  {\bf 87} (2001) 081601.

\bibitem{visc}
  H.~Heiselberg,
  Phys.\ Rev.\ D {\bf 49} (1994) 4739;
  G.~Baym, H.~Monien, C.~J.~Pethick and D.~G.~Ravenhall,
  Phys.\ Rev.\ Lett.\  {\bf 64} (1990) 1867.
%
\bibitem{Teaney:2000cw}
D.~Teaney, J.~Lauret and E.~V.~Shuryak,
Phys.\ Rev.\ Lett.\  {\bf 86} (2001) 4783.
%
\bibitem{Huovinen:2001cy}
P.~Huovinen, P.~F.~Kolb, U.~W.~Heinz, P.~V.~Ruuskanen and S.~A.~Voloshin,
Phys.\ Lett.\ B {\bf 503} (2001) 58.
%
\bibitem{Kolb:2001qz}
P.~F.~Kolb, U.~W.~Heinz, P.~Huovinen, K.~J.~Eskola and K.~Tuominen,
Nucl.\ Phys.\ A {\bf 696} (2001) 197.
%
\bibitem{Hirano:2002ds}
T.~Hirano and K.~Tsuda,
Phys.\ Rev.\ C {\bf 66}, 054905 (2002).
%
\bibitem{Kolb:2002ve}
P.~F.~Kolb and R.~Rapp,
Phys.\ Rev.\ C {\bf 67}, 044903 (2003).
%
\bibitem{Ackermann:2000tr}
K.~H.~Ackermann {\it et al.}  [STAR Collaboration],
Phys.\ Rev.\ Lett.\  {\bf 86} (2001) 402.
%
\bibitem{Adcox:2002ms}
K.~Adcox {\it et al.}  [PHENIX Collaboration],
Phys.\ Rev.\ Lett.\  {\bf 89} (2002) 212301.
%
\bibitem{Adler:2003kt}
S.~S.~Adler {\it et al.}  [PHENIX Collaboration],
Phys.\ Rev.\ Lett.\  {\bf 91} (2003) 182301.
%
\bibitem{Adler:2003cb}
S.~S.~Adler {\it et al.}  [PHENIX Collaboration],
Phys.\ Rev.\ C {\bf 69} (2004) 034909.
%
\bibitem{Adams:2003xp}
J.~Adams {\it et al.}  [STAR Collaboration],
Phys.\ Rev.\ Lett.\  {\bf 92} (2004) 112301.
%
\bibitem{Heinz:2004pj}
  U.~W.~Heinz,
  AIP Conf.\ Proc.\  {\bf 739} (2005) 163.

\bibitem{Baier:2002bt}
  R.~Baier, A.~H.~Mueller, D.~Schiff and D.~T.~Son,
  Phys.\ Lett.\ B {\bf 539} (2002) 46.
%
\bibitem{Molnar:2001ux}
  D.~Molnar and M.~Gyulassy,
  Nucl.\ Phys.\ A {\bf 697} (2002) 495
  [Erratum-ibid.\ A {\bf 703} (2002) 893].


\bibitem{Shuryak:2004cy}
  E.~V.~Shuryak,
  Nucl.\ Phys.\ A {\bf 750} (2005) 64.
  
\bibitem{Lee:2005gw}
  T.~D.~Lee,
  Nucl.\ Phys.\ A {\bf 750} (2005) 1.

\bibitem{Gyulassy:2004zy}
  M.~Gyulassy and L.~McLerran,
  Nucl.\ Phys.\ A {\bf 750} (2005) 30.
%
\bibitem{Heinz:2005zg}
  U.~W.~Heinz,
  arXiv:nucl-th/0512051.
%
\bibitem{Muller:2006ee}
  B.~Muller and J.~L.~Nagle,
  arXiv:nucl-th/0602029.
%
\bibitem{Blau:2005pk}
  S.~K.~Blau,
  Phys.\ Today {\bf 58N5} (2005) 23.
%
\bibitem{Mrowczynski:1996vh}
  S.~Mrowczynski,
  Phys.\ Lett.\ B {\bf 393} (1997) 26.

\bibitem{Arnold:2004ti}
  P.~Arnold, J.~Lenaghan, G.~D.~Moore and L.~G.~Yaffe,
  Phys.\ Rev.\ Lett.\  {\bf 94} (2005) 072302.

\bibitem{Rebhan:2004ur}
  A.~Rebhan, P.~Romatschke and M.~Strickland,
  Phys.\ Rev.\ Lett.\  {\bf 94} (2005) 102303.

\bibitem{Dumitru:2005gp}
  A.~Dumitru and Y.~Nara,
  Phys.\ Lett.\ B {\bf 621} (2005) 89.

\bibitem{Romatschke:2005pm}
  P.~Romatschke and R.~Venugopalan,
  Phys.\ Rev.\ Lett.\ {\bf 96}, (2006) 062302.
%
\bibitem{Hirano:2005xf}
  T.~Hirano, U.~W.~Heinz, D.~Kharzeev, R.~Lacey and Y.~Nara,
  arXiv:nucl-th/0511046.
%
\bibitem{Bhalerao:2005mm}
  R.~S.~Bhalerao, J.~P.~Blaizot, N.~Borghini and J.~Y.~Ollitrault,
  Phys.\ Lett.\ B {\bf 627} (2005) 49.
  %
\bibitem{IS}
W.~Israel, Ann.Phys. {\bf 100} (1976) 310;
W.~Israel and J.M.~Stewart, Phys. Lett. {\bf 58A} (1976)  213;
W.~Israel and J.M.~Stewart, Ann.Phys. {\bf 118}, (1979) 341.
%
\bibitem{Muronga}
  A.~Muronga,
  Phys.\ Rev.\ Lett.\  {\bf 88} (2002) 062302
  [Erratum-ibid.\  {\bf 89} (2002) 159901];
  Phys.\ Rev.\ C {\bf 69} (2004) 034903.
%
\bibitem{Muronga:2004sf}
  A.~Muronga and D.~H.~Rischke,
  arXiv:nucl-th/0407114.
%
\bibitem{Heinz:2005bw}
  A.~K.~Chaudhuri and U.~W.~Heinz,
  arXiv:nucl-th/0504022;
  U.~W.~Heinz, H.~Song and A.~K.~Chaudhuri,
  arXiv:nucl-th/0510014.
%
%
%
\bibitem{Raju1}
  M.~Prakash, M.~Prakash, R.~Venugopalan and G.~M.~Welke,
  Phys.\ Rev.\ Lett.\  {\bf 70} (1993) 1228
  [Nucl.\ Phys.\ A {\bf 566} (1994) 403c].
\bibitem{Raju2}
  M.~Prakash, M.~Prakash, R.~Venugopalan and G.~Welke,
  Phys.\ Rept.\  {\bf 227} (1993) 321.
%
%
\bibitem{Hisc}
W.A.~Hiscock and L.~Lindblom, Phys.\ Rev.\ D {\bf 31}, 725 (1985).
%
\bibitem{CooperFrye}
F.~Cooper and G.~Frye, Phys.\ Rev.\ D {\bf 10}, 186 (1974).
%
\bibitem{Rischke:1998fq}
  D.~H.~Rischke,
  arXiv:nucl-th/9809044,
  \emph{Hadrons in Dense Matter and
Hadrosynthesis}, ed. by J.Cleymans, H.B.~Geyer and F.G.~Scholz, 
Springer Lecture Notes in Physics {\bf 516}, 21 (1999).
%
\bibitem{Bjorken}
J.~Bjorken, Phys.\ Rev.\ D {\bf 27}, 140 (1983).
%
\bibitem{Kouno}
  H.~Kouno, M.~Maruyama, F.~Takagi and K.~Saito,
  Phys.\ Rev.\ D {\bf 41} (1990) 2903.
%
\bibitem{Lallouet}
  Y.~Lallouet, D.~Davesne and C.~Pujol,
  Phys.\ Rev.\ C {\bf 67} (2003) 057901.

\bibitem{Baym}
  G.~Baym,
  Nucl.\ Phys.\ A {\bf 418} (1984) 525c.


%
\bibitem{Teaney}
  D.~Teaney,
  Phys.\ Rev.\ C {\bf 68} (2003) 034913.
  %
\bibitem{Schnedermann}
  E.~Schnedermann, J.~Sollfrank and U.~W.~Heinz,
  Phys.\ Rev.\ C {\bf 48} (1993) 2462.


\bibitem{Ouerdane:2002gm}
  D.~Ouerdane  [BRAHMS Collaboration],
  Nucl.\ Phys.\ A {\bf 715} (2003) 478.

\bibitem{Wosiek:2002ur}
  B.~Wosiek {\it et al.}  [PHOBOS Collaboration],
  Nucl.\ Phys.\ A {\bf 715} (2003) 510.




\bibitem{relax}
  G.~Baym,
  Phys.\ Lett.\ B {\bf 138} (1984) 18;
A.~Hosoya and K.~Kajantie, Nucl.\ Phys.\ B {\bf 250} (1985) 666.

\bibitem{Heiselberg}
  H.~Heiselberg and X.~N.~Wang,
  Nucl.\ Phys.\ B {\bf 462} (1996) 389.


\bibitem{LeBellac}
M.~LeBellac, ``Thermal Field Theory'', Cambridge University Press
(1996)


  
\end{thebibliography}
\end{document}